\documentclass{article}
\usepackage{amsmath}
\usepackage{amstext}
\usepackage{amsfonts}
\usepackage{amssymb,amscd}
\usepackage{amsopn}
\usepackage{graphicx,color}
\usepackage{cite}
\usepackage{bm,bbm}
\usepackage{epsfig} 

\begin{document}
\newcommand{\ors}[1]{{\color{red} #1}}
\newcommand{\comment}[1]{{\color{blue} [#1]}}
\vspace{0.3cm}

\begin{center}

{\Large Three favorite dimensions of Andrzej Kossakowski: \\
   along his path to a scientific discovery}


\vspace{0.3cm}

{\large Karol \.Zyczkowski} 
\medskip

Faculty of Physics, Astronomy and Applied Computer Science, 
Jagiellonian University, Cracow \\
 and \\
 Center for Theoretical Physics, Polish Academy of Sciences, Warsaw


 \ March 19, 2022

\end{center}
\vspace{13mm}
{\bf Abstract}. 
Some achievements of the late Andrzej Kossakowski in the field 
of statistical physics and quantum theory are presented. 
We recall his attempt to find an analytical solution of the 3-dimensional
Ising model and present a problem concerning the set of positive maps.
\vspace{5mm}

\bigskip

The late professor Andrzej Kossakowski (1938 -- 2021)
was a truly multidimensional character: a representative of a
Polish noble family\footnote{It is not difficult to find on the internet the names of his forefathers:
Andrzej was a son of Marian, a grandchild of Miko{\l}aj, a great-grandchild of Walery,
who was a son of J{\'o}zef,   a grandchild of Micha{\l} and a great-grandchild
of Antoni Kossakowski (1735-1798).},
an extremely kind gentleman, an accomplished scientist and teacher.
However, in this note I shall concentrate
on dimension three, which played a prominent role in his research. 

The most well known results of Andrzej
concerning the theory of open quantum systems were established in seventies.
The first decisive steps in the direction of establishing general rules governing the time evolution of quantum systems interacting with an environment
were obtained in a single-author 1972 paper \cite{Ko72}.
 The complete solution of this problem
was published in a 1976 article written jointly with
Gorini and Sudarshan \cite{GKS76}. 
This single paper, cited more than 3500 times in the literature,
is nowadays considered, together with an independent, 
parallel paper of Lindblad \cite{Li76},
as the most important generalization  \cite{CP17}
 of the von Neumann equation  governing the time evolution of a density matrix
in a closed physical system. 
Although in the early eighties, these fundamental results 
were not yet very well-known and appreciated by the world-wide community, 
prof. Kossakowski was already considered as a key figure
in Polish mathematical physics.

\begin{figure}[h]
    \center{
	\includegraphics[angle=0,width=0.65\columnwidth]{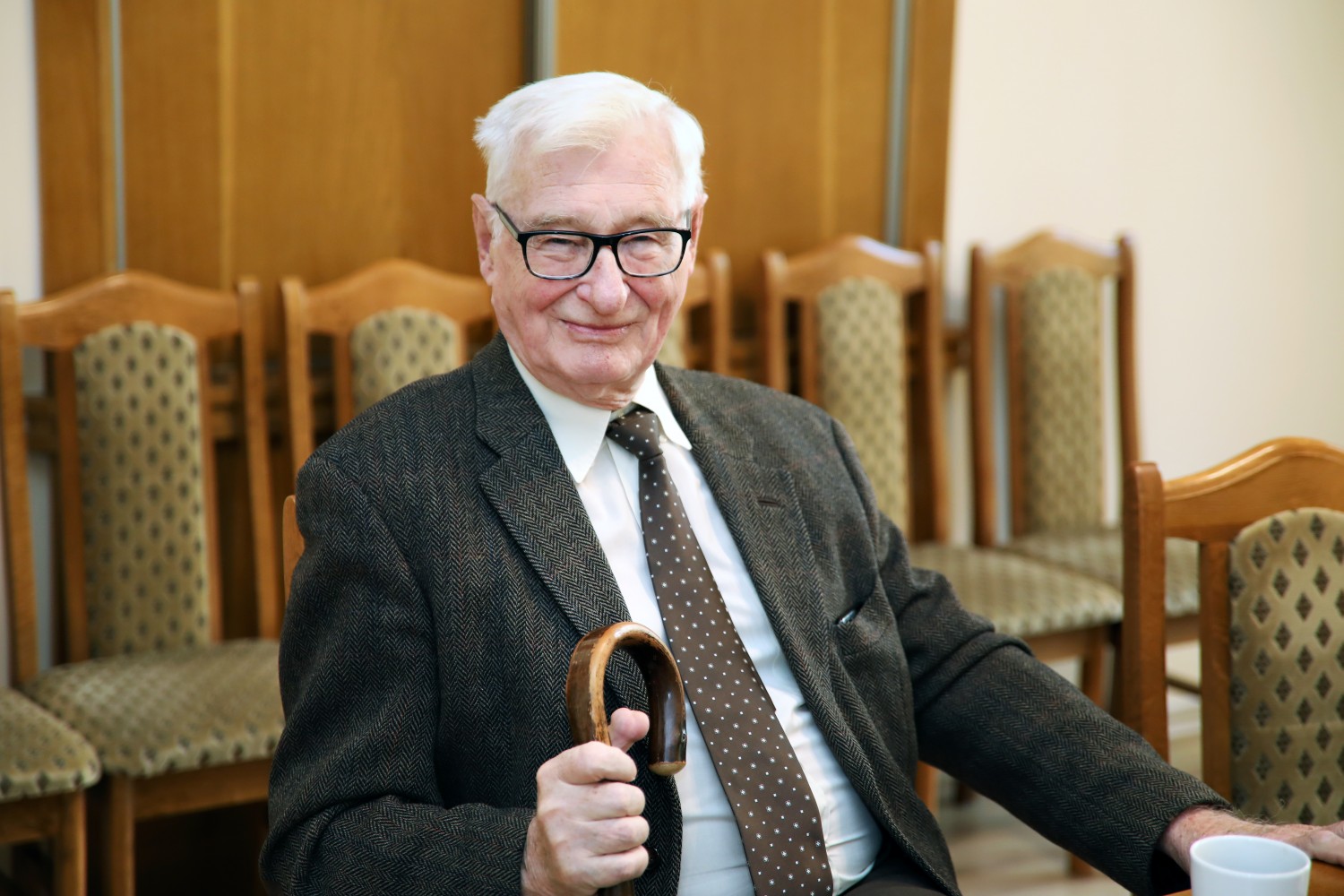}}
	\caption{The late professor Andrzej Kossakowski (1938 -- 2021).}
	\label{Koss}
\end{figure}

\medskip

As I completed my Ph.D thesis in summer 1987,
 my supervisor, Prof. Jan Mostowski from Warsaw
told me briefly:
"Prof. Kossakowski will serve as a referee of your Thesis.
You just go to Toru{\'n}, meet prof. Kossakowski at the Copernicus University
and present what you have achieved to him personally."

In those days it was not possible to arrange a meeting by e-mail
and all contacts required more careful planning.
As I arrived at the Institute of Physics in Toru{\'n}, I met prof. Kossakowski, 
who invited me to his office and said:
"I have already gone through your thesis and I appreciate it."
The thesis was devoted to a model of strong field ionisation due to the
emergence of chaotic dynamics and a quantum version
of the kicked rotator model,
but it contained mostly numerical results, and I was rather
nervous that the referee will say that this is not good enough.
Despite this Andrzej was very polite to me
and after a brief discussion on quantum chaos he changed the topic
and started to talk about the 3D Ising model. 
As I could hardly remember the standard 2D solution of Onsager,
I was shy and embarrassed that my ignorance will soon be revealed. 
However, Andrzej decided not to ask any more questions
but to present his story to a young Ph.D. student:
\medskip 

{\it Alea iacta est} -- {\sl The dice are cast. Now I cannot change anything, 
as I need to go forward in the direction chosen.
It remains just to finish the calculations I started some time ago. 
After a week or two I will eventually realise,
whether the method I have chosen is the right one and 
I will obtain the exact solution concerning the free energy
in the 3D Ising model.
If the answer is positive, it will be a miracle.
If not, I will not be able to do anything more. We shall see.}

\medskip

Andrzej gave me some more insight on the technical details of the 3D Ising problem
and his attempts to solve it. Even though I could 
not understand much, I was highly impressed by the way Andrzej described his results.
Furthermore, I was pleased to be allowed to witness his formidable struggle
to arrive at a groundbreaking result in theoretical physics.

Two months later Andrzej came to Cracow for my Ph.D. defense at the Jagiellonian University.
The defense went fine so I could invite my supervisor and the referees for dinner.
After the dessert I became brave enough to ask him, if there any news about the Ising model.
Andrzej answered something like this:
{\sl The technique used was simply not good enough to provide the desired 
solution of the 3D model.}
But he did not look very upset about it.
And added: {\sl We just need to find another method to solve this problem...}
Even though Andrzej did not succeed in finding an analytical solution 
of the 3D Ising model, he obtained a closed-form approximation for the
free energy in this model published in 1992 in the very first volume of
{\sl Open Systems \& Information Dynamics} \cite{Ko92}.
Later on he continued his research in this direction \cite{Ko94,Ko94b}.

\begin{figure}[h]
    \center{
	\includegraphics[angle=0,width=0.45\columnwidth]{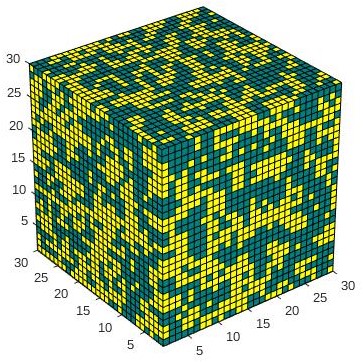}}
	\caption{Exemplary configuration of a $3$D Ising model \cite{Ising}
	 for a cube of size $n=30$ with parameters: exchange integral $J=1$,
	     no external field, $h=0$,
	    and temperature $T=2$. }
	\label{ising}
\end{figure}

\medskip

For the last thirty years I have had the pleasure of visiting Toru{\'n}
 on a regular basis to take part 
in Symposia of Mathematical Physics, initiated by late Professor R.S. Ingarden
and later run by Andrzej Jamio{\l}kowski, Dariusz Chru{\'s}ci{\'n}ski and Mi{\l}osz Michalski.
Each time I visited Prof. Kossakowski in his office at the Copernicus University just to talk to him and 
to learn what he was working on. 
After several years it is easy for me to recall inspiring discussions
we had on the structure of the set of density matrices of an arbitrary finite size \cite{KK05}, on entanglement witnesses \cite{CKS09,CKMM10},
and on non-Markovian quantum dynamics \cite{CK10,CKR11,CK12},
which strongly influenced my own research for the coming decade.
I admired his influential book on Information Dynamics and Open Systems
written jointly with Ingarden and Ohya \cite{IKO97}, which is
directly related to the scope of the present journal.

Our most recent discussions concerned the structure of the set of
positive maps. Due to the  St{\o}rmer--Woronowicz theorem it is known  \cite{St63,Wo76}
that any positive one-qubit map
is decomposable, so it can be represented 
as a convex combination of a completely positive maps and a completely co-positive map.
However, already for $N=3$ there exist non-decomposable positive maps \cite{Ch75},
so the structure of the set of positive maps becomes more intricate.
Even though Andrzej worked on this problem himself \cite{CK09a,CK09b,Ko03},
he also directly encouraged me to work on it.
I believe he was driven by pure curiosity,
as any detailed characterisation of this set is still missing.
As the question, which attracted so much of Andrzej's attention,
remains open, let us conclude this note by formulating it explicitly.
It is convenient to impose the trace preserving condition,
which implies that the set of such maps acting on quantum states of size $N$
forms a convex body
in  $N^4-N^2$ dimensions. Although Andrzej was interested
in the general case of an arbitrary $N$,
we shall state it here for the smallest dimension
for which the problem remains open. Incidentally,
this number coincides with the dimension of the Ising model analyzed. 
\medskip

{\bf Kossakowski problem.} {\sl Characterize the set of positive, trace preserving
   maps acting on quantum states of dimension $N=3$.}

\begin{figure}[h]
    \center{
	\includegraphics[angle=0,width=0.95\columnwidth]{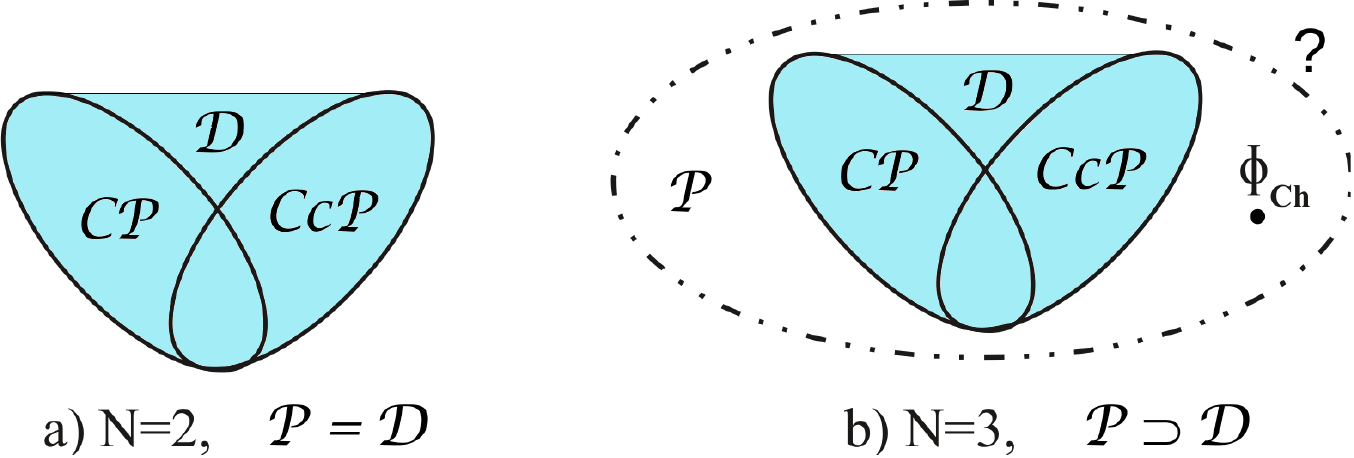}}
	\caption{For $N=2$ the set of positive maps ${\cal P}_2$ 
	   is equal to the set of decomposibe maps:
	     ${\cal D}={\rm conv}\bigl( {\cal CP} \cup {\cal C}c{\cal P}\bigr)$
	        -- see panel a).  As for $N=3$   
	     there exists a non-decomposible map $\Phi_{\rm Ch}$ of Choi,
	      see panel b),
	    the question concerning the structure of the set ${\cal P}_3$ remains open.}
	\label{Decomp}
\end{figure}

\medskip

I am pretty sure, Andrzej would be pleased to learn
how this $72$--dimensional set of positive, trace preserving
 maps of a qutrit might look...

\end{document}